\documentclass[12pt]{iopart}

\usepackage{amsfonts}

\begin{document}

\title{Quantum State Tomography Using Successive Measurements}
\author{Amir Kalev$^1$ and Pier A. Mello$^2$}
\address{$^1$Centre for Quantum Technologies, National University of Singapore, Singapore 117543}

\address{$^2$Departamento de Sistemas Complejos, Instituto de
F\'{\i}sica, Universidad Nacional Aut\'{o}noma de M\'{e}xico,
M\'{e}xico, D.F. C.P. 04510}

\ead{cqtamirk@nus.edu.sg, mello@fisica.unam.mx}

\begin{abstract}
We describe a quantum state tomography scheme which is applicable to 
a system described in a Hilbert space of arbitrary finite dimensionality
and is constructed from sequences of two measurements.
The scheme consists of measuring the various pairs of projectors onto two bases 
--which have no mutually orthogonal vectors--, the two members of each pair being measured in succession.
We show that this scheme implies measuring the joint quasi-probability of  any pair of non-degenerate observables having the two bases as their respective eigenbases. The model Hamiltonian underlying the scheme makes use of two meters initially prepared in an arbitrary given quantum state, following the ideas that were introduced by von Neumann in his theory of measurement.
\end{abstract}


\vspace{5cm}

\maketitle

\section{Introduction}
\label{introduction}

The fundamental problem of inferring the initially unknown state of a quantum system from a set of measured quantities, i.e., the problem of quantum-state reconstruction, goes back to the early days of quantum mechanics, when it was known as Pauli problem \cite{pauli}. 
Aside from its fundamental importance, the ability to reconstruct and characterize quantum states has implications in various areas of technology and information sciences.
Since a large amount of theoretical and experimental work has been devoted to this problem, we refer the reader to a representative number of contributions only and references therein \cite{newton,wootters,weigert92,amiet_weigert99,amiet_weigert_L99,james-kwiat-munro-white,englert,filippov10-2}. 

The main goal in formulating a tomographic scheme is to identify a set of measurable quantities that gives complete information about the state of the system. It was shown \cite{wootters} that a complete set of mutually unbiased bases (MUB) is of this kind and could be used for state tomography with high efficiency. However, the construction of a complete set of MUB is known for powers of primes dimensions only \cite{durt10}. A possible alternative to MUB for state tomography is to use what are known as symmetric informationally complete probability-operator measurements (SIC POMs) \cite{englert}.
These kind of measurements are known to exist  in all dimensions $d\leq 45$ (with high numerical precision) \cite{renes04}. 
So far, however, all experiments and even proposals for experiments implementing  SIC POMs have been limited to the very basic quantum system, the two-level system (qubit) \cite{ling06}. This is, in part, due to the fact that there is no systematic procedure for implementing SIC POMs in higher dimensions, in a simple experimental setup. 

In this contribution we identify sets of observables which, when measured in succession, provide complete information about the state of a quantum system described in a $d$-dimensional Hilbert space. 
The tomographic scheme consists of measuring the various pairs of projectors onto two bases which have no mutually orthogonal vectors, the two members of each pair being measured in succession.
We formulate the scheme using the von Neumann (vNM) model for measurements extended to two successive measurements. We find that in the formulation the notion of finite-dimensional quasi-probability distribution (the analogue of the phase-space quasi-probability distribution) appears in a natural manner.

The paper is organized as follows.
In the next section we describe the dynamics of successive measurements of two arbitrary observables as an extension of the vNM model, assuming that the initial state of the two meters is described by an arbitrary density operator. 
In Sec.~\ref{tomography-scheme} we use this formalism to discuss the reconstruction scheme based on successive measurements. 
This procedure generalizes that of  Ref. \cite{johansen-mello} to the more general case discussed in the present article. In Sec.~\ref{transform} we show that the notion of quasi-probability distribution and generalized transform of observables that was introduced in Ref.~\cite{johansen-mello} can  also be applied to the present general case.  Finally, we present our conclusions in Sec.~\ref{conclusions}.

\section{The successive measurements of two observables}
\label{succ-meas-2obs}

Consider the successive measurement of two observables: $\hat{A}$ is measured first and $\hat{B}$ later. 
For this purpose we generalize the standard vNM \cite{von_neumann,aharonov_et_al}, assuming two meters $M_i$
(with canonical momentum and coordinate operators $\hat{P}_i$, $\hat{Q}_i$, $i=1,2$)
which interact successively with the system of interest according to the interaction
\begin{equation}
\hat{V} (t) 
= \epsilon_1 g_1 (t-t_1) \hat{A} \hat{P}_1
+ \epsilon_2 g_2 (t-t_2) \hat{B} \hat{P}_2  \; ,
\label{V(t)}
\end{equation}
with $0 < t_1 < t_2$.
The functions $g_1$ and $g_2$ are normalized to 1 and have a (non-overlapping) compact support around the corresponding interaction times $t_1$ and $t_2$. 
We denote by $\epsilon_i$ $(i=1,2)$ the strength of the interaction between the system and meter $M_i$. 

Before the first interaction,  
the system and the two meters are described by the density operators $\rho_s$, $\rho_{M_1}$ and $\rho_{M_2}$, respectively, and the combined system by their tensor product 
$\rho =  \rho_{s} \rho_{M_1} \rho_{M_2}$.
After the second interaction, the combined system is described by 
\cite{johansen-mello}
\begin{eqnarray}
&\rho^{(\hat{B} \leftarrow \hat{A})} 
= \sum_{nn'mm'}
\mathbb{P}_{b_{m}} \mathbb{P}_{a_{n}}  \rho_{s} \; \mathbb{P}_{a_{n'}} \mathbb{P}_{b_{m'}}
\nonumber \\
&\hspace{1cm}\cdot 
e^{-i\epsilon_1 a_n \hat{P}_1} \rho_{M_1}
e^{i\epsilon_1 a_{n'}\hat{P}_1} 
\cdot
e^{-i\epsilon_2 b_m \hat{P}_2}  \rho_{M_2}
e^{i\epsilon_2 b_{m'} \hat{P}_2},
\label{rho t>t2}
\end{eqnarray}
where $(\hat{B} \leftarrow \hat{A})$ indicates that $\hat{A}$ has been measured first and $\hat{B}$ later.
Of course, the whole process could be considered as one global measurement, leading to the density operator (\ref{rho t>t2}).
Here, $\hat{A}$ and $\hat{B}$ are expressed in their spectral representation
\numparts
\begin{eqnarray}
\hat{A} = \sum_n a_n \mathbb{P}_{a_n},  
\label{A-spectral repres} \\
\hat{B} = \sum_m b_m \mathbb{P}_{b_m}.
\label{B-spectral repres}
\end{eqnarray}
\endnumparts
The eigenprojectors $\mathbb{P}_{a_n}$ and $\mathbb{P}_{b_m}$ correspond to the possibly degenerate
eigenvalues $a_n$ of $\hat{A}$  and $b_m$ of 
$\hat{B}$, respectively;
they satisfy the orthogonality and completeness relations 
\numparts
\begin{eqnarray}
\mathbb{P}_{a_n} \mathbb{P}_{a_{n'}}  
= \delta_{n n'}  \mathbb{P}_{a_n},
\label{orthonormality}
\\
\sum_n \mathbb{P}_{a_n} 
= \mathbf{1}, 
\label{completeness}
\end{eqnarray}
\endnumparts
and similarly for $\hat{B}$.

We now seek information on the system by observing the position-position correlation function of the two meters, $\langle \hat{Q}_1\hat{Q}_2\rangle$, where the average is taken over the state of Eq.~(\ref{rho t>t2}).
Under the assumption that the average initial position of the two meters vanishes, we obtain
\numparts
\begin{eqnarray}
\frac{1}{\epsilon_1 \epsilon_2}
\langle \hat{Q}_1 \hat{Q}_2 \rangle\
^{(\hat{B} \leftarrow \hat{A})}
= \Re  \sum_{nm} a_{n} b_{m}
W^{(\hat{B} \leftarrow \hat{A})}_{b_{m}a_{n}}(\epsilon_1),
\label{<Q1Q2> BA}
\end{eqnarray}
where $\Re$ stands for the `real part' and we have defined
\begin{eqnarray}
W^{(\hat{B} \leftarrow \hat{A})}_{b_{m}a_{n}}(\epsilon_1)
=\sum_{n'}
\lambda(\epsilon_1(a_n - a_n')) \;
{\rm tr}(\rho_{s}\mathbb{P}_{a_{n'}} \mathbb{P}_{b_{m}} \mathbb{P}_{a_{n}})\; .
\label{W(b,a) BA}
\end{eqnarray}
\endnumparts
The function $\lambda(\epsilon_1(a_n - a_n'))$ is explicitly given in \ref{lambda,lambda_tilde}.

Consider again the same Hamiltonian of Eq.~(\ref{V(t)}), but suppose that at the end of the measurement procedure 
(i.e., for $t>t_2$) we observe the momentum $\hat{P}_1$ of the first meter (instead of its position) and the position $\hat{Q}_2$ of the second meter. 
The resulting momentum-position correlation function  is
\numparts
\begin{eqnarray}
\frac1{\epsilon_1 \epsilon_2}\langle \hat{P}_1 \hat{Q}_2 \rangle
 ^{(\hat{B} \leftarrow \hat{A})}
=2\sigma^2_{P_1}
\Im \sum_{nm}  a_{n} b_{m}
\widetilde{W}^{(\hat{B} \leftarrow \hat{A})}_{b_{m}a_{n}}(\epsilon_1)
\label{<P1Q2> BA}
\end{eqnarray}
where $\Im$ stands for `imaginary part', $\sigma^2_{P_1}$ is the second moment of $\hat{P}_1$ in the initial meter state,  and  we have defined
\begin{eqnarray}
\widetilde{W}^{(\hat{B} \leftarrow \hat{A})}_{b_{m}a_{n}}(\epsilon_1)
&=&\sum_{n'}
\tilde{\lambda}(\epsilon_1(a_n - a_n'))
{\rm tr}(\rho_{s}\mathbb{P}_{a_{n'}} \mathbb{P}_{b_{m}} \mathbb{P}_{a_{n}}).
\hspace{10mm}
\label{W_tilde BA}
\end{eqnarray}
\endnumparts
The function $\tilde{\lambda}(\epsilon_1(a_n - a_n'))$ can be found in  \ref{lambda,lambda_tilde} as well.
The quantity $W^{(\hat{B} \leftarrow \hat{A})}_{b_{m}a_{n}}(\epsilon_1)$ of Eq.~(\ref{W(b,a) BA}) is in general different from  $\widetilde{W}^{(\hat{B} \leftarrow \hat{A})}_{b_{m}a_{n}}(\epsilon_1)$ 
of Eq.~(\ref{W_tilde BA}), because $\lambda$ is in general different from $\tilde{\lambda}$.
For pure Gaussian states of the meters, $\lambda=\tilde{\lambda}$ and therefore 
$W=\widetilde{W}$. This was the starting point for the reconstruction scheme presented in \cite{johansen-mello}. In what follows we formulate a tomographic scheme in which this equality may not hold.

\section{State tomography scheme}
\label{tomography-scheme}

We now use the above formalism to describe a state tomography scheme.
For this purpose we consider a $d$-dimensional Hilbert space and two orthonormal bases, whose vectors are denoted by $|k \rangle$ and $|\mu \rangle$, respectively, with $k, \mu=1,\ldots, d$. Latin letters will be used to denote the first basis while Greek letters will be used for the second basis.
We assume the two bases to be mutually non-orthogonal, i.e.,
$\langle k| \mu\rangle \neq 0$, $\forall k,\mu$.
This last condition implies that the two bases have 
no common eigenvectors and are said to be complementary \cite{beltrametti}.
The condition that the two bases have no common eigenvectors is equivalent to requiring that two observables having these bases as their eigenbases should never possess simultaneous definite values.
An example of mutually non-orthogonal bases are two bases which are related by the Fourier transform.

We now consider the following meters-system interaction
\begin{equation}
\hat{V} (t) 
= \epsilon_1 g_1 (t-t_1) \mathbb{P}_{k} \hat{P}_1
+ \epsilon_2 g_2 (t-t_2) \mathbb{P}_{\mu} \hat{P}_2 \; ,
\label{V(t)proj}
\end{equation}
with $0 < t_1 < t_2$.
Here, $\mathbb{P}_{k} = |k\rangle \langle k|$ and
$\mathbb{P}_{\mu} = |\mu\rangle \langle \mu|$ are rank-one projectors onto the $k$- and $\mu$-state of the first and second basis, respectively.
The observable $\hat{A}$ appearing in the interaction of Eq.~(\ref{V(t)}) is replaced here by the projector $\mathbb{P}_{k}$ 
and $\hat{B}$ by the projector $\mathbb{P}_{\mu}$. 
Being projectors, these observables possess two eigenvalues: $0$ and $1$.
We denote by $\tau$ and $\sigma$ the eigenvalues of $\mathbb{P}_{k}$ and  $\mathbb{P}_{\mu}$, respectively, and the corresponding eigenprojectors by
$(\mathbb{P}_{k})_{\tau}$ and $(\mathbb{P}_{\mu})_{\sigma}$.
We have
\numparts
\begin{eqnarray}
(\mathbb{P}_{k})_{1}
= \mathbb{P}_{k} \\
\label{P1}
(\mathbb{P}_{k})_{0} 
= \mathbf{1} - \mathbb{P}_{k} = \sum_{k'(\neq k)} \mathbb{P}_{k'},
\label{P0}
\end{eqnarray}
\label{P1,0}
\endnumparts
and similarly for $(\mathbb{P}_{\mu})_{1}$ and $(\mathbb{P}_{\mu})_{0}$.

In the present case, Eq.~(\ref{<Q1Q2> BA}) for the meters position-position correlation function gives
\numparts
\begin{eqnarray}
\frac{1}{\epsilon_1 \epsilon_2}
\langle \hat{Q}_1 \hat{Q}_2 \rangle
^{(
\mathbb{P}_{\mu} \leftarrow \mathbb{P}_{k}
)}
&= \Re \sum_{\tau, \sigma = 0}^1 \tau \sigma \;
W^{( \mathbb{P}_{\mu} \leftarrow \mathbb{P}_{k})}
_{\sigma \tau} (\epsilon_1)
\label{<Q1Q2> P P a} 
\\
&= 
\Re W^{( \mathbb{P}_{\mu} \leftarrow \mathbb{P}_{k})}_{1 1} (\epsilon_1).
\label{<Q1Q2> P P b}
\end{eqnarray}
\label{<Q1Q2> P P}
\endnumparts
In Eq. (\ref{<Q1Q2> P P a}), 
$W^{( \mathbb{P}_{\mu} \leftarrow \mathbb{P}_{k})}
_{\sigma \tau} (\epsilon_1)$
is the particular case of the quantity
$W^{(\hat{B} \leftarrow \hat{A})}_{b_{m}a_{n}}(\epsilon_1)$
of Eq. (\ref{W(b,a) BA}) 
when $\hat{A}$, $\hat{B}$, $a_n$, and $b_m$ are replaced by $\mathbb{P}_{k}$, $\mathbb{P}_{\mu}$, $\tau$, and $\sigma$, respectively, i.e.,
\begin{equation}
W^{(\mathbb{P}_{\mu} \leftarrow \mathbb{P}_{k})}
_{\sigma \tau} (\epsilon_1)
=\sum_{\tau'=0}^1 \lambda(\epsilon_1(\tau-\tau'))
{\rm tr}\left[\rho_s  (\mathbb{P}_{k})_{\tau'} (\mathbb{P}_{\mu})_{\sigma} 
(\mathbb{P}_{k})_{\tau} \right] \; ,
\end{equation}
and, in particular,
\begin{equation}
W^{( \mathbb{P}_{\mu} \leftarrow \mathbb{P}_{k})}_{1 1} (\epsilon_1)
=
{\rm tr}(\rho_{s}\mathbb{P}_{k}\mathbb{P}_{\mu} \mathbb{P}_{k})
+\lambda(\epsilon_1) 
\sum_{k' (\neq k)} {\rm tr} (\rho_{s}
\mathbb{P}_{k'}\mathbb{P}_{\mu} \mathbb{P}_{k}) \; ,
\label{W11 1}
\end{equation}
where we have used Eq. (\ref{lambda(0)=1}) of  \ref{lambda,lambda_tilde}.
Now we invert Eq.~(\ref{W11 1}) to obtain $\rho_s$.
This inversion, which was briefly indicated in Ref. \cite{johansen-mello},
is valid under the more general situation contemplated here.
To see this, we first write Eq. (\ref{W11 1}) as
\begin{equation}
W^{( \mathbb{P}_{\mu} \leftarrow \mathbb{P}_{k})}_{1 1} (\epsilon_1)
=\left[ \sum_{k'} G_{k' k}(\epsilon_1)
\langle k|\rho_s|k' \rangle  \langle k'|\mu \rangle \right]
\langle \mu| k\rangle \; ,
\label{W11 2}
\end{equation}
where
\begin{equation}
G_{k' k}(\epsilon_1) 
= \delta_{k' k} + \lambda(\epsilon_1) (1 - \delta_{k' k}).
\label{G}
\end{equation}
From Eq.~(\ref{W11 2}) we obtain
\begin{eqnarray}
\sum_{\mu} 
\frac{W^{( \mathbb{P}_{\mu} \leftarrow \mathbb{P}_{k})}_{1 1} (\epsilon_1)}{\langle \mu| k\rangle}
\langle \mu| k'\rangle
=  G_{k' k}(\epsilon_1) \langle k|\rho_s|k' \rangle  ,
\label{W11 3}
\end{eqnarray}
so that
\numparts
\begin{eqnarray}
\langle k|  \rho_{s}   | k' \rangle
= \sum_{\mu}
\frac{W^{( \mathbb{P}_{\mu} \leftarrow \mathbb{P}_{k})}
_{1 1} (\epsilon_1)}
{G_{k'k}(\epsilon_1)} \cdot 
\frac{\langle \mu|k'\rangle}
{\langle \mu|k\rangle}  ,
\label{rho elements from W(b,a)}
\end{eqnarray}
and hence
\begin{eqnarray}
\rho_{s}  
= \sum_{k,\mu}
W^{( \mathbb{P}_{\mu} \leftarrow \mathbb{P}_{k})}_{1 1} (\epsilon_1)
\left[ 
|k\rangle\left(1-\frac1{\lambda(\epsilon_1)}\right)\langle k|+|k\rangle\frac1{\lambda(\epsilon_1)\langle \mu|k\rangle}\langle \mu|
\right].
\nonumber \\
\label{rho from W(b,a)}
\end{eqnarray}
\endnumparts
Eqs. (\ref{rho elements from W(b,a)}) and (\ref{rho from W(b,a)}) are the main result of this paper. 
They imply that $\forall k,\mu$,
$W^{( \mathbb{P}_{\mu} \leftarrow \mathbb{P}_{k})}_{1 1}$
of Eq. (\ref{W11 1}), just as 
$\rho_s$,
contains all the information about the state of the system. Therefore, inferring  $W^{( \mathbb{P}_{\mu} \leftarrow \mathbb{P}_{k})}
_{1 1}\;\forall k,\mu$ from the measurement outcomes is equivalent to the reconstruction of  $\rho_s$. Our aim is to show that the measured quantities, the position-position and momentum-position correlation functions, are informationally complete: 
that is, one can reconstruct $W^{( \mathbb{P}_{\mu} \leftarrow \mathbb{P}_{k})}
_{1 1}$  from these quantities.
Note that neither the strength of the second interaction nor the state of the second meter enter Eq. (\ref{rho from W(b,a)}).

We notice that the full complex quantity  $W^{( \mathbb{P}_{\mu} \leftarrow \mathbb{P}_{k})}_{1 1}$ is needed for tomography.
From the position-position correlation function of Eq.~(\ref{<Q1Q2> P P b}) we \textit{directly} extract the real part of  
$W^{(\mathbb{P}_{\mu} \leftarrow \mathbb{P}_{k})}_{11}$. 
To find the imaginary part of $W_{11}$ we measure the momentum-position correlation function, which in the present case is given by
\begin{equation}
\frac{1}{\epsilon_1 \epsilon_2}
\langle \hat{P}_1 \hat{Q}_2 \rangle
^{(
\mathbb{P}_{\mu} \leftarrow \mathbb{P}_{k}
)}
= 2 \sigma_{P_1}^2 \Im \widetilde{W}^{( \mathbb{P}_{\mu} \leftarrow \mathbb{P}_{k})}
_{1 1} (\epsilon_1),
\label{<P1Q2> P P}
\end{equation}
where 
\begin{equation}
\widetilde W^{( \mathbb{P}_{\mu} \leftarrow \mathbb{P}_{k})}_{1 1} (\epsilon_1)
= {\rm tr} (\rho_{s}\mathbb{P}_{k}\mathbb{P}_{\mu} \mathbb{P}_{k})
+\tilde\lambda(\epsilon_1) \sum_{k'(\neq k)} {\rm tr} (\rho_{s}
\mathbb{P}_{k'}\mathbb{P}_{\mu} \mathbb{P}_{k})\; .
\label{tildeW P P 0}
\end{equation}
We note in passing that this equation can be inverted to write $\rho_{s}$ in terms of $\widetilde W$, resulting in Eq.~(\ref{rho elements from W(b,a)}) with $W$ and $\lambda$ are replaced by $\widetilde W$ and $\tilde\lambda$. Though the function $\widetilde{W}$ is in general not equal to the function $W$, we prove in \ref{W11} that it contains all the information about the imaginary part of $W^{( \mathbb{P}_{\mu} \leftarrow \mathbb{P}_{k})}_{1 1}$, and therefore enables a complete state reconstruction. This completes our procedure.

At first glance it seems that, in a $d$-dimensional Hilbert space, the present scheme for state reconstruction requires the measurement of 
the meters position-position and momentum-position correlations for the $d^2$ successive measurements of projectors, $\mathbb{P}_{k}$ followed by $\mathbb{P}_{\mu}$, giving $2d^2$ different measurements altogether.
However, Hermiticity and the unit value of the trace of the density matrix $\rho_s$ impose $d^2+1$ restrictions among its matrix elements, so that $\rho_s$
can be expressed in terms of $d^2-1$ independent parameters. 
These restrictions eventually imply that only $d^2-1$ of these correlations are actually independent and thus 
the measurement of only $d^2-1$ correlations is required. \ref{tomography_N=2} gives an application of the above formalism to the case of a two-dimensional Hilbert space, and shows explicitly how the matrix elements of $\rho_{s}$ can be expressed in terms of $d^2-1=3$ independent measurable correlations.

Finally, we close this section with the following remarks. In the infinitely-strong coupling limit, $\epsilon_1 \to \infty$, $\lambda(\epsilon_1)$ vanishes, and $W_{11}$ contains information only about the diagonal elements of $\rho_s$, as can be seen from Eq.~(\ref{W11 1}). 
In the other extreme of weak coupling, 
in particular in the limit when $\epsilon_1 \to 0$, 
$W_{11}$ contains the full information  about the state of the system, cf. Eq.~(\ref{W11 1}). This limit was the result  presented in Ref. \cite{johansen07}. 
Therefore, to reconstruct a quantum state using the successive-measurement scheme it is better to perform a measurement with a weak coupling to the first meter rather than one with a strong coupling.

\section{A quasi-distribution and a generalized transform of observables}
\label{transform}

From a conceptual point of view,
one attractive feature of the present approach is related to the quantities
$W^{( \mathbb{P}_{\mu} \leftarrow \mathbb{P}_{k})}_{1 1} (\epsilon_1)$
that enter the state reconstruction formula, Eqs. (\ref{rho elements from W(b,a)}) and (\ref{rho from W(b,a)}). 
This quantity can be interpreted as a ``joint quasi-probability distribution'' 
in the following sense.
Let $\hat{O}$ be an observable associated with a $d$-dimensional quantum system.
Making use of Eq. (\ref{rho elements from W(b,a)}) we can express its expectation value as
\begin{equation}
{\rm tr}(\rho_s \hat{O})
=\sum_{k k'}
\langle k|  \rho_{s}   | k' \rangle
\langle k'| \hat{O} | k \rangle
=\sum_{k\mu} 
W^{( \mathbb{P}_{\mu} \leftarrow \mathbb{P}_{k})}_{1 1} (\epsilon_1)
\;
O(\mu,k,\epsilon_1),
\label{<O> 2}
\end{equation}
where we have defined the ``transform" of the operator $\hat{O}$ as
\begin{eqnarray}
O(\mu,k,\epsilon_1)
&= \sum_{k'} 
\frac{\langle \mu|k'\rangle}{\langle \mu|k\rangle}
\frac{\langle k'| \hat{O} | k \rangle}{G_{kk'}(\epsilon_1)}  \nonumber\\
&=\left(1-\frac1{\lambda(\epsilon_1)}\right)\langle k|\hat{O} | k \rangle+\frac1{\lambda(\epsilon_1)}\frac{\langle \mu|\hat{O} | k\rangle}{\langle \mu|k\rangle} \; .
\label{Omn}
\end{eqnarray}
Eq. (\ref{<O> 2}) has the structure of a number of transforms found in the literature, that express the quantum mechanical expectation value of an observable in terms of its transform and a quasi-probability distribution.
For example, the Wigner transform of an observable and the Wigner function of a state are defined in the phase space $(q,p)$ of the system, $q$ and $p$ labelling 
the states of the coordinate and momentum bases, 
respectively. 
In the present case, the transform (\ref{Omn}) of the observable is defined for the pair of variables $(\mu, k)$, $\mu$ and $k$ labelling the states of 
each of the two bases.
As Eq. (\ref{<O> 2}) shows, it is the quantity 
$W^{( \mathbb{P}_{\mu} \leftarrow \mathbb{P}_{k})}_{1 1} (\epsilon_1)$ 
which plays the role of the quasi-probability for the system state
$\hat{\rho}_s$, and is also defined for the pair of variables 
$(\mu, k)$.
It can be thought of as the joint quasi-probability of two non-degenerate observables, with the two bases being their respective eigenbases. 
Since any pair of mutually non-orthogonal bases can be used, we have a whole family of transforms that can be employed to retrieve the state.

In the literature it has been discussed how Wigner's function can be considered as a representation of a quantum state 
(Ref. \cite{schleich}, Chs. 3 and 4), in the sense that i) it allows retrieving the density operator, and ii) any quantum-mechanical expectation value can be evaluated from it.
Similarly, and for the same reasons, in the present context the quasi-probability 
$W^{( \mathbb{P}_{\mu} \leftarrow \mathbb{P}_{k})}_{1 1} (\epsilon_1)$
can also be considered as a representation of a quantum state.

\section{Conclusions}
\label{conclusions}

We discussed successive measurements as an alternative approach to realize informationally complete measurements on quantum systems. 
Here we considered a particular Hamiltonian model for successive measurements that involves the system proper and both meters, and is an extension to two meters of vNM of measurement.

In the approach presented in this paper we considered, in the $d$-dimensional Hilbert space of the system, 
two complete, orthonormal bases, assumed to be 
mutually non-orthogonal, also called complementary.
The observables needed for the present scheme are the projectors onto the basis vectors of each one of these two bases.  We then showed that the set of all pairs of successive measurements of such projectors, one for each basis, allows the complete retrieval of the system state (Eqs. (\ref{rho elements from W(b,a)}), (\ref{rho from W(b,a)})).
We proved that the scheme can be formulated for arbitrary states of the meters, and for an arbitrary strength of the meter-system interaction.

We showed that this procedure can be interpreted as measuring the joint quasi-probability of pairs of non-commuting observables, in a 
way similar to the state reconstruction based on measuring the quasi-probability in phase space provided by the Wigner transform of the state.

As a final note we wish to point out two possible extensions that might be of interest. 
One is the state tomography of a continuous-variable system using successive measurements. 
The second generalization concerns state tomography 
when both the system and the meters are described in a finite-dimensional Hilbert space. 

\section{Acknowledgments}
The Centre for Quantum Technologies (CQT) is a Research Centre of Excellence funded by the Ministry of Education and the National Research Foundation of Singapore. 
P.A.M acknowledges financial support from CQT, Singapore, and 
CONACyT, Mexico, under Contract 79501.
The authors are grateful to B.-G. Englert and C. A. M\"uller for enlightening discussions. 

\appendix

\section{The quantities $\lambda(\beta)$ of Eq. (\ref{W(b,a) BA}) and $\tilde{\lambda}(\beta)$ of Eq. (\ref{W_tilde BA})}
\label{lambda,lambda_tilde}

We are using the notation $\langle \cdots \rangle_{M_1}
= \tr (\rho_{M_1} \cdots)$. Let us define
\numparts
\begin{eqnarray}
g(\beta)
&= \langle e^{-i \beta \hat{P}_1}\rangle_{M_1},
\label{g}
\\
h(\beta) 
&= \frac{1}{\beta}\langle 
e^{-i \frac{\beta}{2}\hat{P}_1} \hat{Q}_1 e^{-i \frac{\beta}{2}\hat{P}_1}\rangle_{M_1},
\label{h}
\end{eqnarray}
where $\beta=\epsilon_1(a_n - a_{n'})$. Since
$\langle \hat{Q}_1 \rangle_{M_1}=0$, Eq.~(\ref{h}) is well defined when $\beta \to 0$. The function
$\lambda(\beta)$ is defined as
\begin{eqnarray}
\lambda(\beta)
&= g(\beta) + 2 h(\beta).
\label{lambda}
\end{eqnarray}
\label{lambda,g,h}
\endnumparts
The functions $g(\beta)$ and $h(\beta)$ satisfy the properties
\numparts
\begin{eqnarray}
g^*(\beta) &= g(-\beta),
\hspace{8mm} g(0) = 1,
\label{g(0)}
\\
h^*(\beta)& = -h(-\beta), \hspace{5mm} 
h(0) = -\frac{i}{2}
\langle  \hat{Q}_1\hat{P}_1 + \hat{P}_1\hat{Q}_1 \rangle_{M_1} =0 ,
\label{h(0)}
\end{eqnarray}
and therefore 
\begin{eqnarray}
\lambda(0)=1.
\label{lambda(0)=1}
\end{eqnarray}
\endnumparts
In writing Eq.~(\ref{h(0)}) we have assumed the natural condition that the current density at the point $Q_1$ for the first meter prior to the measurement
\begin{equation}
J(Q_1)
= \frac{1}{2m_1} \langle \mathbb{P}_{Q_1} \hat{P_1} + \hat{P_1} \mathbb{P}_{Q_1} \rangle_{M_1}
\label{J(Q)}
\end{equation}
vanishes $\forall Q_1$ ($m_1$ being the mass of the first meter, and $\mathbb{P}_{Q_1}=|Q_1\rangle\langle Q_1|$). Thus
\numparts
\begin{eqnarray}
\int J(Q_1) dQ_1 &= \frac{1}{m_1} \langle \hat{P}_1 \rangle_{M_1} = 0,
\label{<P1>}
\\
\int Q_1 J(Q_1) dQ_1 
&= \frac{1}{2 m_1} \langle \hat{Q}_1 \hat{P}_1 + \hat{P}_1 \hat{Q}_1 \rangle_{M_1} = 0,
\label{<PQ+QP>}
\end{eqnarray}
\label{<P>,<QP+PQ>}
\endnumparts
from which Eq. (\ref{h(0)}) follows. 
This assumption does not affect the generality of our results and could be lifted in a straightforward manner.

The function $\tilde{\lambda}(\epsilon_1(a_n - a_n'))$ is defined as follows.
We first define 
\numparts
\begin{eqnarray}
\bar{\lambda}(\beta)= \frac{1}{\beta} \frac{\partial g(\beta)}{\partial \beta},
\label{lambda_bar}
\end{eqnarray}
where $g(\beta)$ is given in Eq. (\ref{g}). We are assuming that $\langle \hat{P}_1 \rangle_{M_1} = 0$, so that $\bar{\lambda}(\beta)$ is well defined when $\beta \to 0$: indeed, we find the series expansion
\begin{eqnarray}
\bar{\lambda}(\beta)
= -\sigma_{P_1}^2 + i \frac{\beta}{2!}\langle \hat{P}_1^3 \rangle_{M_1} + \cdots.
\label{lambda_bar-series}
\end{eqnarray}
\label{lambda_bar-and-series}
\endnumparts
We then define $\tilde{\lambda}(\beta)$ as
\numparts
\begin{eqnarray}
\tilde{\lambda}(\beta)
&=& \frac{\bar{\lambda}(\beta)}{\bar{\lambda}(0)} 
\label{lambda_tilde}
\\
&=& 1 -  i \frac{\beta}{2 \sigma_{P_1}^2}\langle \hat{P}_1^3 \rangle_{M_1} + \cdots.
\label{lambda_tilde_series}
\end{eqnarray}
\label{lambda_tilde-and-series}
\endnumparts

\section{Construction of $W_{11}$ from $\Re W_{11}$ and $\Im \tilde{W}_{11}$}
\label{W11}

If we write 
\numparts
\begin{eqnarray}
W^{(\mathbb{P}_{\mu} \leftarrow \mathbb{P}_{k})}_{1 1} 
= x_{\mu k} + i y_{\mu k}, 
\label{x,y} \\
\widetilde{W}^{(\mathbb{P}_{\mu} \leftarrow \mathbb{P}_{k})}_{1 1} 
= \tilde{x}_{\mu k} + i  \tilde{y}_{\mu k},
\label{x,y,xtilde,ytilde}
\end{eqnarray}
\endnumparts
the correlation functions,
Eqs. (\ref{<Q1Q2> P P b}) and (\ref{<P1Q2> P P}), become
\numparts
\begin{eqnarray}
\frac
{\langle \hat{Q}_1 \hat{Q}_2 \rangle
^{(
\mathbb{P}_{\mu} \leftarrow \mathbb{P}_{k}
)}}
{\epsilon_1 \epsilon_2}
= x_{\mu k}, 
\label{<Q1Q2>=x}         \\
\frac
{\langle \hat{P}_1 \hat{Q}_2 \rangle
^{(
\mathbb{P}_{\mu} \leftarrow \mathbb{P}_{k}
)}}
{\epsilon_1 \epsilon_2}
= 2 \sigma_{P_1}^2 \tilde{y}_{\mu k}.
\label{<P1Q2>=ytilde}
\end{eqnarray}
\endnumparts
Our aim is to express $y_{\mu k}$ in terms of the measured quantities.
We go back to the expressions (\ref{W11 1}) and (\ref{tildeW P P 0})
for 
$W^{( \mathbb{P}_{\mu} \leftarrow \mathbb{P}_{k})}_{1 1} (\epsilon_1)$ and
$\widetilde{W}^{( \mathbb{P}_{\mu} \leftarrow \mathbb{P}_{k})}_{1 1} (\epsilon_1)$.
The quantities $\lambda(\epsilon_1)$, $\tilde{\lambda}(\epsilon_1)$  are known if the state of the measuring apparatus $M_1$ is known 
[see Eqs. (\ref{g}), (\ref{h}), (\ref{lambda})] and (\ref{lambda_tilde})];
we write
\numparts
\begin{eqnarray}
\lambda(\epsilon_1) 
= \lambda_r(\epsilon_1) + i \lambda_i(\epsilon_1), 
\label{lambda r,i}        \\ 
\tilde{\lambda}(\epsilon_1) 
= \tilde{\lambda}_r(\epsilon_1) + i \tilde{\lambda}_i(\epsilon_1).
\label{lambda_tilde r,i}
\end{eqnarray}
\endnumparts
On the other hand, the traces appearing in Eqs. (\ref{W11 1}) and 
(\ref{tildeW P P 0}) are unknown; we write them as
\numparts
\begin{eqnarray}
 \tr (\rho_{s}\mathbb{P}_{k}\mathbb{P}_{\mu} \mathbb{P}_{k})
&= r^0_{\mu k} 
\label{r0}
\\
\sum_{k'(\neq k)} \tr (\rho_{s}
\mathbb{P}_{k'}\mathbb{P}_{\mu} \mathbb{P}_{k})
&= r_{\mu k}+is_{\mu k} .
\label{r,s}
\end{eqnarray}
\label{r0,r,s}
\endnumparts
Using Eq. (\ref{rho elements from W(b,a)}), $r^0_{\mu k}$ of Eq. (\ref{r0}) can be written in terms of measured quantities only, as
\begin{equation}
r^0_{\mu k}
=  |\langle k | \mu \rangle |^2 
\langle k |\rho_s | k \rangle
= |\langle k | \mu \rangle |^2
\sum_{\mu'} x_{\mu' k} .
\label{r0 1}
\end{equation}
We introduce the definitions (\ref{x,y},{\em b}), 
(\ref{lambda r,i},{\em b}), (\ref{r0},{\em b}) and the result  
(\ref{r0 1}) in Eqs.~(\ref{W11 1}) and (\ref{tildeW P P 0}) and write the latter as
\numparts
\begin{eqnarray}
x_{\mu k}
&=& |\langle k | \mu \rangle |^2
\sum_{\mu'} x_{\mu' k} + \lambda_r r_{\mu k} -\lambda_i  s_{\mu k} ,
\label{x-mu-k}    
\\
y_{\mu k}
&=& \lambda_i 
r_{\mu k}  
+ \lambda_r
s_{\mu k}, 
   \\
\tilde{y}_{\mu k}
&=& \tilde{\lambda}_i 
r_{\mu k}  
+  \tilde{\lambda}_r
s_{\mu k}.
\end{eqnarray}
\endnumparts
For every pair of indices $\mu, k$ we now have a system of three linear equations in the three unknowns $r_{\mu k}$, $s_{\mu k}$ and $y_{\mu k}$, which can thus be expressed in terms of the measured quantities 
$\tilde{y}_{\mu k}$ and the $x_{m k}$
of Eq. (\ref{<Q1Q2>=x},{\em b}).
The result for $y_{\mu k}$ is
\begin{equation}
y_{\mu k}
= \frac{\Im \{\lambda(\epsilon_1)\tilde\lambda^\ast(\epsilon_1)\}}
{\Re \{\lambda(\epsilon_1)\tilde\lambda^\ast(\epsilon_1)\}}
\;\Big(x_{\mu k}-|\langle k|\mu\rangle|^2 
\sum_{\mu'} x_{\mu' k}\Big)
+\frac{|\lambda(\epsilon_1)|^2}{\Re \{\lambda(\epsilon_1)\tilde\lambda^\ast(\epsilon_1)\}}\;\tilde{y}_{\mu k} .
\label{y}
\end{equation}

We have thus achieved our goal of expressing 
$W^{( \mathbb{P}_{\mu} \leftarrow \mathbb{P}_{k})}_{1 1} (\epsilon_1)$, 
and hence  $\rho_s$ of Eq.~(\ref{rho from W(b,a)}),
in terms of the measured correlations of Eqs.~(\ref{<Q1Q2>=x},{\em b}).

\section{State reconstruction for a two-level system}
\label{tomography_N=2}

As an illustrative example of the general formulation of Sec. \ref{tomography-scheme}, we consider the state reconstruction of a two-level system.

We take the projectors $\mathbb{P}_{k}$ with $k=0,1$ and $\mathbb{P}_{\mu}$ with $\mu=\pm$ as projectors onto the eigenstates of the Pauli operators $\sigma_z$ and $\sigma_x$, respectively. 
The measured quantities are the correlation functions $\langle \hat{Q}_1 \hat{Q}_2 \rangle^{(\mathbb{P}_{\mu} \leftarrow \mathbb{P}_{k})}$ and  $\langle \hat{P}_1 \hat{Q}_2 \rangle^{(\mathbb{P}_{\mu} \leftarrow \mathbb{P}_{k})}$. In the case of spin-1/2 particles, the interaction of Eq.~(\ref{V(t)proj}) could be realized by cascading two Stern-Gerlach apparatuses, equipped with additional constant electric fields, and one measures the particle position-position and momentum-position correlations in the $\hat{z}$ and $\hat{x}$ directions. 

For simplicity of the presentation of the example, we choose the case in which $\lambda(\epsilon_1)$ and $\tilde{\lambda}(\epsilon_1)$ defined in \ref{lambda,lambda_tilde} are real. 
It can be shown that this can be achieved if the original state of the first meter is an arbitrary mixture of pure states which, in the coordinate representation, are real and of definite parity. 
As a result, from Eq. (\ref{y}) we find that the $y_{\mu k}$ are related to the measurable 
$\tilde{y}_{\mu k}$ as
\begin{equation}
y_{\mu k} =  \frac{\lambda(\epsilon_1)}{\tilde{\lambda}(\epsilon_1)} \tilde{y}_{\mu k} \; .
\label{y vs ytilde}
\end{equation} 
The $\langle k|\rho_{s}|k'\rangle \equiv \rho_{kk'}$ elements of the density matrix,
Eq. (\ref{rho elements from W(b,a)}), can be expressed in terms of the $8$ real quantities 
$x_{\mu k}$, $y_{\mu k}$ of Eq. (\ref{x,y}) 
(or $x_{\mu k}$, $\tilde{y}_{\mu k}$, using (\ref{y vs ytilde})).The conditions of Hermiticity and unit trace of $\rho_{k k'}$ give 5 relations among the 8 correlations $x_{\mu k}$ and $\tilde{y}_{\mu k}$, so that only 3 of them are independent.
One can choose 
\numparts
\begin{eqnarray}
x_{+0}
= \frac
{\langle \hat{Q}_1 \hat{Q}_2 \rangle
^{( \mathbb{P}_{+} \leftarrow \mathbb{P}_{0})}}
{\epsilon_1 \epsilon_2} \; ,
\label{x+0}  \\
x_{-0}
= \frac
{\langle \hat{Q}_1 \hat{Q}_2 \rangle
^{( \mathbb{P}_{-} \leftarrow \mathbb{P}_{0})}}
{\epsilon_1 \epsilon_2} \; ,
\label{x-0} \\
\tilde{y}_{-0}
= \frac
{\langle \hat{P}_1 \hat{Q}_2 \rangle
^{( \mathbb{P}_{-} \leftarrow \mathbb{P}_{0})}}
{2 \sigma_{P_1}^2 \epsilon_1 \epsilon_2} \; ,
\label{y-0}
\end{eqnarray}
\endnumparts 
as the 3 independent correlations, and one finds that the remaining 5 can be expressed in terms of them as
\numparts
\begin{eqnarray}
x_{+1} = \frac1{2} - x_{-0} , 
\label{dep_correls_vs_indep a} \\
x_{-1} = \frac1{2}- x_{+0} , 
\label{dep_correls_vs_indep b} \\
\tilde{y}_{+0} = - \tilde{y}_{-0},
\label{dep_correls_vs_indep c} \\
\tilde{y}_{+1} = \tilde{y}_{-0},
\label{dep_correls_vs_indep d} \\
\tilde{y}_{-1} = - \tilde{y}_{-0} \; .
\label{dep_correls_vs_indep e}
\end{eqnarray}
\endnumparts 
Finally, the $\rho_{k k'}$ matrix elements can be written in terms of
measured quantities as
\numparts
\begin{eqnarray}
\rho_{00} = x_{+0} + x_{-0} ,
\label{rho_00} \\
\rho_{11} = 1 - x_{+0} - x_{-0} ,
\label{rho_11} \\
\rho_{01} =
\frac{1}{\lambda(\epsilon_1)}
(x_{+0}  - x_{-0})
- i \; \frac{2}{\tilde{\lambda}(\epsilon_1)} \; \tilde{y}_{-0} \; ,
\label{rho_01} \\
\rho_{10} = \rho_{01}^{*} \; .
\end{eqnarray}
\endnumparts
%

\end{document}